\documentstyle[twocolumn,twoside,psfig]{article}
\newcommand{\hb}{\\ \hspace*{2ex}}

\begin{document}
\title{COMMENTS ON ENVIRONMENTAL EFFECTS IN THE ORIGIN OF ANGULAR MOMENTA IN GALAXIES}
\author{
Elena Panko${^1}$, Paulina Pajowska ${^2}$, W{\l}odzimierz God{\l}owski${^3}$,
Piotr Flin $^{4}$\\[5mm]
\begin{tabular}{l}
$^1$ Nikolaev National University, Kalinenkov Astronomical Observatory, \hb Nikolskaya, 24, Nikolaev, 54030, Ukraine {\em panko.elena@gmail.com}\\
$^2$ Opolski University, Institute of Physics,\hb ul. Oleska  48,
45-052 Opole, Poland {\em paoletta@interia.pl}\\
$^3$ Opolski University, Institute of Physics,\hb ul. Oleska  48,
45-052 Opole, Poland {\em godlowski@uni.opole.pl}\\
$^4$ Jan Kochanowski University, Institute of Physics,\hb 25-406 Kielce, ul.
Swietokrzyska 15, Poland {\em sfflin@cyf-kr.edu.pl}\\[2mm]
\end{tabular}
}
\date{}
\maketitle


ABSTRACT. We examine the orientations of galaxies in 43 rich Abell
galaxy clusters belonging to superclusters and containing at least
100 members in the considered area as a function of supercluster
multiplicity. It is found that the orientation of galaxies in the
analyzed clusters is not random and the alignment decreases with
supercluster richness, although the effect is statistically
significant only for azimuthal angles. The dependence of galaxy
alignment on cluster location inside or outside a supercluster and
on supercluster multiplicity clearly shows the importance of
environmental effects on the origin of galaxy angular momenta. The
comparison with alignment of galaxies in a sample of rich Abell
clusters not belonging to superclusters is made too.\\[1mm]

{\bf Key words}: galaxies, angular momenta.\\[1mm]

{\bf 1. Introduction}\\[1mm]

One of the most important but unsolved until now problems in
modern extragalactic astronomy and cosmology is the origin of
large scale structures. At present the $\Lambda$CDM model is
commonly accepted as the basis by which cosmic structures were
born. In the model the Universe is considered to be spatially
flat, homogeneous and isotropic at an appropriate scale. However,
the dimension of that scale is changing with the growth in our
knowledge of the Universe. In addition, it is also commonly
accepted that currently observed structures originated from nearly
isotropic distributions in the early Universe. The departure from
isotropy, as estimated by the CMBR is on the order of
$\delta\rho/\rho\simeq 10^{-5}$. About half a century ago the main
problems were connected with the types of perturbations, their
amplitude, and scale (mass or length). In the $\Lambda$CDM model
the structures were formed in the primordial, adiabatic, nearly
scale invariant, Gaussian, random fluctuations.

Numerous different theories of galaxy origins predict various
means by which galaxies gained angular momentum (Peebles 1969,
Zeldovich 1970, Sunyaev \& Zeldovich 1972, Doroshkevich 1973,
Shandarin 1974, Wesson 1982, Silk \& Efstathiou 1983, Bower et al.
2005). Since different scenarios forecast different distributions
for the angular momenta of galaxies in structures (Peebles 1969,
Doroshkevich 1973, Shandarin 1974, Silk \& Efstathiou 1983,
Catelan \& Theuns 1996, Li 1998, Lee \& Pen 2002, Navarro et al.
2004, Trujillio et al. 2006), testing galaxy orientations
can be used to check the scenarios of galaxy origins. Normally
studies of the orientation of galaxy planes were performed.

God{\l}owski \& Flin (2010) studied the
orientation of galaxy groups in the Local Supercluster, and
found a strong alignment of the major axis of the groups with
directions towards the supercluster center (Virgo cluster)
as well as with the line joining the two brightest galaxies in the
group. The interpretation of these observational features is as
follows. The brightest galaxies (believed to be the most massive
ones) of the group originated first. As a result of gravitational
forces, other galaxies were attracted to them and a filament
was formed at the end.

Similar results were obtained by Paz et al. (2011), where the authors
found a strong alignment between the projected major axis of
group shapes and the surrounding distribution of galaxies to
scales of $30h^{-1} Mpc$. Smargon et al. (2011) searched for two
types of cluster alignments using pairs of clusters: the
alignment between the projected major axes of the clusters
displayed a weak effect up to $20 h^{-1} Mpc$, whereas the alignment
between the major axis of one cluster with the line connecting the
other cluster in the pair displayed a strong alignment on scales up to
$100 h^{-1} Mpc$. Also, a statistically significant anisotropy
for the galaxy groups and cluster orientations for a sample of the
Jagellonian field was noted by Flin \& Vavilova (1996) and
Vavilova (1999).

The other possibility for interpreting the (God{\l}owski \&
Flin, 2010) result is that the galaxies form at a
pre-existing filament. Consistent with that argument are the
results of Jones et al. (2010), who found that
the spins of spiral galaxies located within cosmic web filaments
tend to be aligned along the larger axis of the filament. Jones et
al. (2010) interpreted it as ``fossil'' evidence, indicating that the
action of large scale tidal torques effected the alignment of
galaxies located in cosmic filaments. The relationship between
alignment and the surrounding neighborhood was observed in a study
of orientations in the vicinity of voids by Varela et al. (2011),
a continuation of an earlier study of galaxy orientations in
regions surrounding bubble-like voids (Trujillo et al., 2006).
Varela et al. (2011) found that the observed tendency in the alignment
of galaxies is similar to that observed in numerical
simulations of the distribution of dark matter, i.e. in
distributions of the minor axis of dark matter halos around cosmic
voids, which suggests a possible link to the evolution of both
components.

The large scale distribution is usually known as the ``Cosmic Web.'' In
practice the ``Cosmic Web'' has four components which are: long
filaments, walls, voids, and rich, dense regions {\---} so called
galaxy clusters. Thus, we should investigate the alignment of galaxies
and clusters in such structures as well.

In  God{\l}owski et al. (2010), Paper I hereinafter,
a sample of 247 rich Abell clusters was analyzed. It was found that
the alignment of members in rich structures containing more than 100
galaxies is a function of the group mass, in the sense that the
alignment increases with the richness of the group. In view of
such features, it is interesting to see if clusters
belonging to the larger structures exhibit the same type of
alignment as the entire sample of clusters. For that reason, we
God{\l}owski et al. (2011, Paper II hereinafter) analyzed the
alignment of galaxy cluster members for clusters
belonging to superclusters. The problem was not investigated
previously, although the alignment of galaxies in superclusters has been
investigated many times.

In Paper II the alignment of galaxies in the sample of $43$ rich
Abell galaxy clusters belonging to a supercluster and having at
least 100 members was investigated. It was found that the
orientation of galaxies in the analyzed clusters is not random.
However, significant differences were found with the results
obtained in Paper I, in which an increase of alignment was found
for rich Abell clusters as a function of cluster richness. On the
contrary, other clusters belonging to superclusters do not show
such an effect. In Paper I galaxies in the sample studied were
split into three bins according to supercluster multiplicity. They
were: a subsample of superclusters containing only 4 structures, a
subsample of superclusters containing 5{\---}7 structures, and
finally a subsample of superclusters containing 8{\---}10
clusters. However, because the analysis was based on only 3 bins,
it was difficult to determine the statistical significance of the
results. In the present paper we decided to analyse the
orientation of galaxies in clusters belonging to supercluster in
more detail, without binning on clusters properties such as
richness or BM type. In essence, we used the likelihood of
membership in a supercluster as the parameter which characterizes
each analyzed cluster.
\\[2mm]

{\bf 2. Observational data}\\[1mm]

Input data for the present study made use of the PF Catalogue of
galaxy structures (Panko \& Flin, 2006). That Catalogue was
constructed by finding structures in the Muenster Red Sky Survey
(MRSS) (Ungruhe et al., 2003) in conjunction with the Voronoi
tessellation technique applied to find structures. The MRSS is a
large-sky galaxy catalogue covering an area of about 5000 square
degrees in the southern hemisphere. It is the result of scanning
217 ESO plates, yielding positions, red magnitudes, radii,
ellipticities, and position angles for about 5.5 million galaxies,
and is complete to $r_F=18.3^m$. As a result, there are 6188
galaxy structures called clusters. Structure ellipticities and
position angles were determined by means of the standard
covariance ellipse method. We have selected a sample of 247 very
rich clusters containing at least 100 members each that are
identified with an ACO cluster (Abell et al., 1989) -- see Paper I
for more details. Unfortunately there are not obvious correlation
between "rich" PF clusters and Abbell' richness classes. The PF
catalogue was also used as the basis for supercluster search (see,
for example, Panko, 2011) and 54 superclusters containing at least
4 clusters each were detected. We found that $43$ of a total of
$247$ rich PF clusters belong to superclusters, and they were
chosen for detailed analysis. However, it should be noted that
three clusters, 0347-5571, 2217-5177, and 2234-5249, have two
possible identifications with superclusters of different
multiplicity, so must be counted in two bins, which formally
enlarged our sample to $46$ clusters.
\\[2mm]


\begin{table}[h]
\begin{center}
\scriptsize \caption{The frequency of anisotropy of very rich
clusters located in superclusters.}

\label{tab:t1}
\begin{tabular}{cccc}
\hline \\[1mm]
Multiplicity     &   The angle & The angle     &  The angle\\
                 &       $P$     & $\delta_D$    &    $\eta$ \\
 \hline

   N=4           &      0.84   &          0.74  &      0.84\\
   N=5\---7      &      0.31   &          0.90  &      0.79\\
   N=8\---10     &      0.43   &          0.57  &      0.43\\
\hline
\end{tabular}
\end{center}
\end{table}


\begin{table*}[h]
\begin{center}
\scriptsize
\caption{The theoretical (random) and observational values of statistics.}
\label{tab:t2}
\begin{tabular}{c|cccc|cc}
\hline
\multicolumn{1}{c}{Test}& \multicolumn{4}{c}{Simulations}& \multicolumn{2}{c}{Observations}\\
\multicolumn{1}{c} {}   & \multicolumn{4}{c} {}          & \multicolumn{2}{c}{PA}\\
\hline
&$\bar{x}$&$\sigma(x)$&$\sigma(\bar{x})$&$\sigma(\sigma(x))$&$\bar{x}$&$\sigma(x)$ \\
\hline
 $\chi^2$                       & 34.9592 & 1.2843 & 0.0406 & 0.0287& 38.772& 1.574\\
 $\Delta_{1}/\sigma(\Delta_{1})$&  1.2567 & 0.0983 & 0.0031 & 0.0022&  1.797& 0.148\\
 $\Delta/\sigma(\Delta)$        &  1.8846 & 0.1027 & 0.0032 & 0.0023&  2.339& 0.148\\
 $C$                            & -0.9750 & 0.8593 & 0.0274 & 0.0192&  0.611& 1.120\\
 $\lambda$                      &  0.7729 & 0.0392 & 0.0012 & 0.0009&  0.932& 0.057\\
\hline
\end{tabular}
\end{center}
\end{table*}


\begin{figure*}
\hskip 0.2 cm \psfig{figure=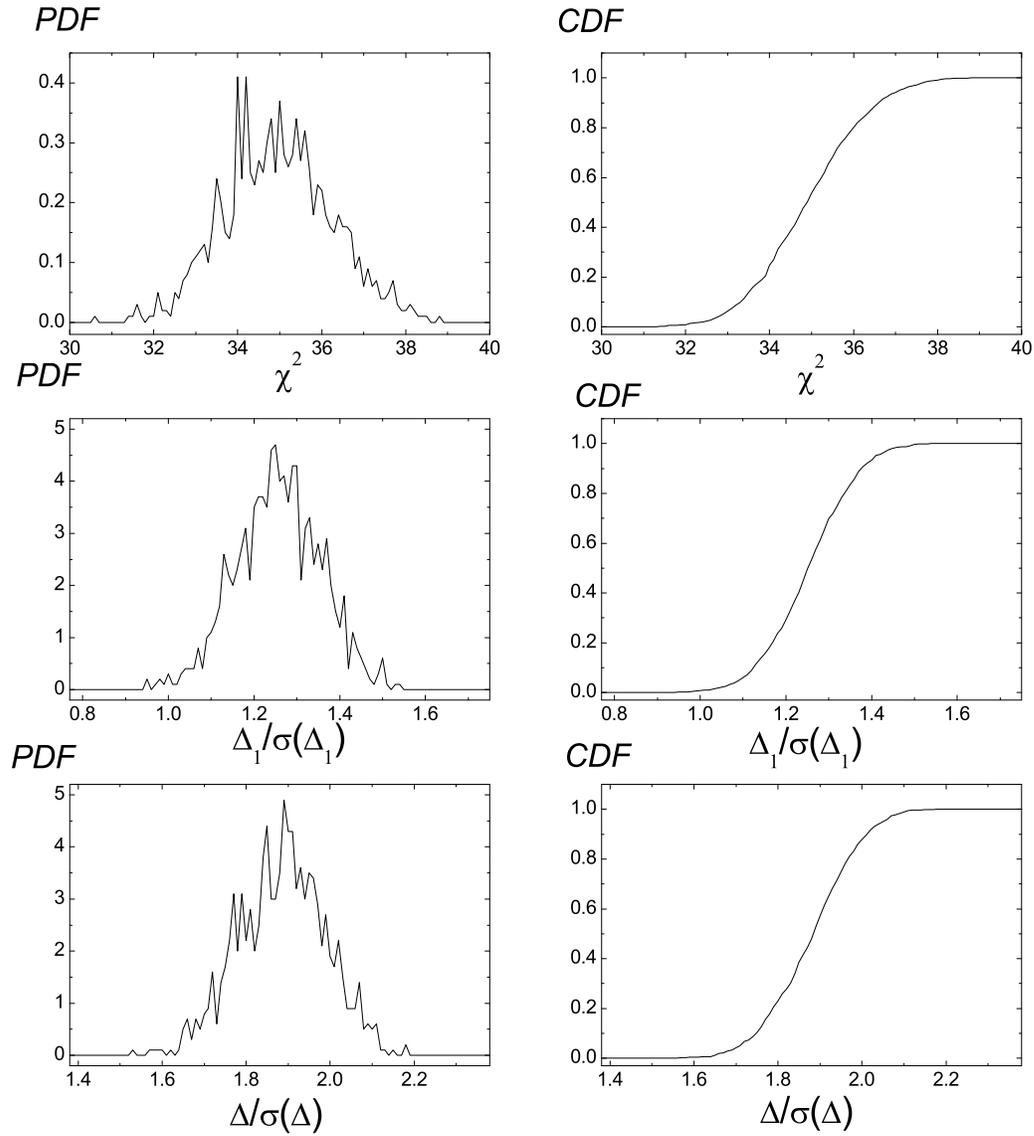,height=6.6in}
\caption{Probability density function (PDF, left panel) and
cumulative distribution function (CDF, right panel) for statistics
$\chi^2$, $\Delta_1/\sigma(\Delta_1)$ and $\Delta/\sigma(\Delta)$.
\label{fig1}}
\end{figure*}


\begin{figure*}
\hskip 0.2cm \psfig{figure=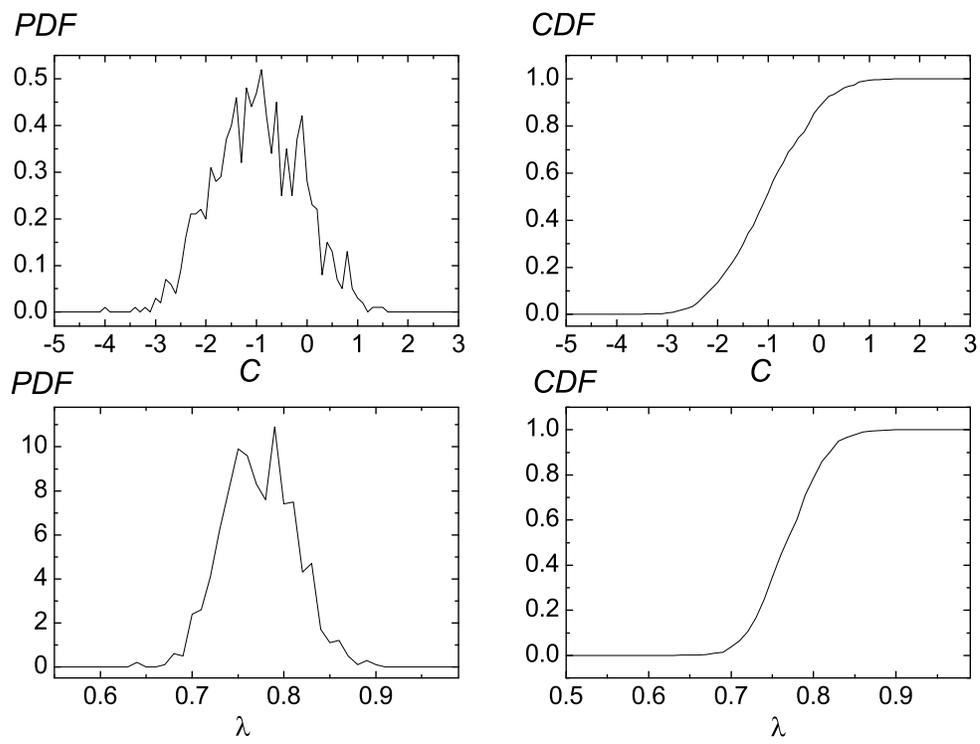,height=4.5in}
\caption{Probability density function (PDF, left panel) and
cumulative distribution function (CDF, right panel) for statistics
$C$ and $\lambda$. \label{fig2}}
\end{figure*}


\begin{table*}[h]
\begin{center}
\scriptsize
\caption{The results of the linear regression analysis: value
of the analyzed statistics as a function of the cluster richness for clusters
belonging to superclusters. \label{tab:t3}}
\begin{tabular}{c|c|c|c|c|c}
\hline
\multicolumn{1}{c}{}&
\multicolumn{1}{c}{$\chi^2$}&
\multicolumn{1}{c}{$\Delta_1/\sigma(\Delta_1)$}&
\multicolumn{1}{c}{$\Delta/\sigma(\Delta)$}&
\multicolumn{1}{c}{$C$}&
\multicolumn{1}{c}{$\lambda$}\\
\hline
angle&$a \pm\sigma(a)$&$a\pm\sigma(a)$&$a\pm\sigma(a)$&$a\pm\sigma(a)$&$a\pm\sigma(a)$\\
\hline
$P$     &$0.028\pm 0.039$&$-.0004\pm 0.0037$&$-.0029\pm 0.0037$&$0.027\pm 0.028$&$0.0004\pm 0.0014$\\
$\delta$&$0.026\pm 0.037$&$-.0008\pm 0.0023$&$0.0020\pm 0.0024$&$0.031\pm 0.032$&$0.0004\pm 0.0011$\\
$\eta$  &$0.125\pm 0.040$&$0.0060\pm 0.0030$&$0.0082\pm 0.0028$&$0.112\pm 0.041$&$0.0018\pm 0.0009$\\
\hline
\end{tabular}
\end{center}
\end{table*}


\begin{table*}[t]
\begin{center}
\scriptsize \caption{The statistical analysis: value of the analyzed
statistics for different supercluster multiplicities.}
\label{tab:t4}
\begin{tabular}{ccccc}
\hline
angle&Test&$N=4$&$N=5-7$&$N=8-10$\\
\hline
$P$           &$\chi^2                       $&$43.30 \pm 2.42$&$  34.99 \pm 2.11$&$  36.65 \pm  1.55$\\
              &$\Delta_{1}/\sigma(\Delta_{1})$&$ 1.99 \pm 0.25$&$   1.50 \pm 0.16$&$   1.89 \pm  0.32$\\
              &$\Delta/\sigma(\Delta)        $&$ 2.57 \pm 0.23$&$   2.08 \pm 0.18$&$   2.42 \pm  0.36$\\
              &$C                            $&$ 2.53 \pm 1.60$&$  -0.99 \pm 1.27$&$  -0.37 \pm  4.05$\\
              &$\lambda                      $&$ 1.01 \pm 0.08$&$   0.81 \pm 0.06$&$   1.01 \pm  0.21$\\
\hline
$\delta_D$&$\chi^2_c                         $&$54.52 \pm10.69$&$ 48.68 \pm 4.57$&$   89.07 \pm 18.56$\\
          &$|\Delta_{11}/\sigma(\Delta_{11})|$&$ 3.13 \pm 0.63$&$  2.76 \pm 0.57$&$    6.41 \pm  0.84$\\
          &$\Delta_{c}/\sigma(\Delta_{c})    $&$ 4.76 \pm 0.72$&$  4.40 \pm 0.34$&$    6.97 \pm  1.09$\\
              &$C_c                          $&$29.73 \pm 9.48$&$ 22.53 \pm 4.46$&$   51.59 \pm 16.36$\\
              &$\lambda_c                    $&$ 2.22 \pm 0.33$&$  2.01 \pm 0.13$&$    3.39 \pm  0.55$\\

\hline
$\eta$       &$\chi^2                        $&$83.60 \pm12.09$&$  95.57 \pm 9.74$&$  43.01 \pm  5.44$\\
             &$\Delta_{1}/\sigma(\Delta_{1}) $&$ 5.43 \pm 0.86$&$   5.83 \pm 0.71$&$   2.02 \pm  0.29$\\
             &$\Delta/\sigma(\Delta)         $&$ 6.30 \pm 0.88$&$   7.31 \pm 0.66$&$   3.19 \pm  0.38$\\
              &$C                            $&$49.31 \pm 12.34$&$ 54.25 \pm 10.50$&$  5.26 \pm  2.93$\\
              &$\lambda                      $&$ 2.24 \pm 0.28$&$   2.37 \pm 0.19$&$   1.07 \pm  0.13$\\
\hline
\end{tabular}
\end{center}
\end{table*}


\begin{table*}
\begin{center}
\scriptsize \caption{The results of linear regression analysis:
value of the analyzed statistics as a function of
supercluster multiplicity. \label{tab:t5}}
\begin{tabular}{c|c|c|c|c|c}
\hline
\multicolumn{1}{c}{}&
\multicolumn{1}{c}{$\chi^2$}&
\multicolumn{1}{c}{$\Delta_1/\sigma(\Delta_1)$}&
\multicolumn{1}{c}{$\Delta/\sigma(\Delta)$}&
\multicolumn{1}{c}{$C$}&
\multicolumn{1}{c}{$\lambda$}\\
\hline
angle&$a \pm\sigma(a)$&$a\pm\sigma(a)$&$a\pm\sigma(a)$&$a\pm\sigma(a)$&$a\pm\sigma(a)$\\
\hline
$P$     &$-1.093\pm 0.772$&$ 0.018\pm 0.074$&$ 0.001\pm 0.074$&$-0.249\pm 0.563$&$ 0.014\pm 0.029$\\
$\delta$&$ 6.839\pm 2.937$&$ 0.527\pm 0.181$&$ 0.435\pm 0.195$&$ 4.282\pm 2.651$&$ 0.206\pm 0.126$\\
$\eta$  &$-7.863\pm 3.553$&$-0.574\pm 0.255$&$-0.558\pm 0.256$&$-7.772\pm 3.668$&$-0.178\pm 0.078$\\
\hline
\end{tabular}
\end{center}
\end{table*}

{\bf 3. Results and Discussion}\\[1mm]

Studies of galaxy alignments are usually done by analyzing
distributions of the angles connected with the orientation of the
galaxy plane; namely the position angle of the great axis of the
galaxy image $P$ and the angles describing the orientation of the
normal to the galaxy plane: $\delta_D$ and $\eta$. The polar angle
$\delta_D$ is the angle between the normal to the galaxy plane and
the main plane of the coordinate system; the azimuthal one $\eta$
is angle between the projection of this normal onto the main plane
and the direction toward the zero initial meridian and positional
angle (see for example Flin \& God{\l}owski 1986, Paper I). In the
present paper, as well as in Paper II, we analyzed the sample of
43 very rich clusters (having 100 and even more members) belonging
to superclusters.

The existence of an alignment for each particular cluster
belonging to our sample was analyzed in Paper I (Table 4). On
that basis it was possible to analyze the frequency of
alignments in our sample of galaxy clusters attributed to
superclusters (Table 1). To first order, the data indicate that
anisotropy decreases with supercluster multiplicity.

It is also possible to analyze the alignment of clusters
belonging to superclusters in more detail. The standard method of
approach for galactic alignments is an analysis of the
distribution of the angles, which provides information connected with
the orientation of galaxies. That approach was proposed by Hawley
\& Peebles (1975), who analyzed the distribution of position
angles using $\chi^2$ testing, Fourier testing, and first autocorrelation
testing. One should note that there were several modifications and
improvements to the original Hawley \& Peebles (1975) method (Flin
\& God{\l}owski 1986, Kindl 1987, God{\l}owski 1993, 1994, Aryal
\& Saurer 2000, God{\l}owski et al., 2010). God{\l}owski (2012)
made a significant improvement to the original Hawley \& Peebles
(1975) method and showed its usefulness in the analysis of
galaxy orientations in clusters. In God{\l}owski
(2012) the mean values of the analyzed statistics were computed. The
null hypothesis $H_0$ assumed that the mean value of the analyzed
statistics was that expected for the case of a random distribution of
analyzed angles. The results were compared with theoretical
predictions as well as with the results obtained from numerical
simulations.

Following the God{\l}owski (2012) method, we analyzed our sample
of $43$ clusters belonging to superclusters. In Paper II we
analyzed only the $\chi^2$ statistic and statistics obtained on
the basis of Fourier testing.

$\chi^2$ statistics was studied as

\begin{equation}
\chi^{2}=\sum\limits_{k=1}^{n}\frac{(N_{k}-N_{0,k})^{2}}{N_{0,k}}
\end{equation}

were $N_{k}$ is the number of galaxies within $k-th$ angular bin
and as $N_{0,k}$ is the expected number of galaxies per bin, n is
the number of bins.

If the theoretical probability function $PF$ is uniform, then
$N_{0,k}$ are equal.

In all applied statistical tests, the entire range of the
investigated $\theta$ angle (as $\theta$ we accept
$\delta_{D}+\pi/2$, $\eta$ or $P$).

If deviation from isotropy is a slowly varying function of the è
angle, one can use the Fourier test (Hawley \& Peebles, 1975):
\begin{equation}
N_{k} = N_{0,k}(1 +
\Delta_{11}\cos2\theta_{k}+\Delta_{21}\sin2\theta_{k})+...
\end{equation}

If the theoretical probability function $PF$ is symmetric with
respect to the value $\theta=\pi/2$, we obtain the following
expressions for the Fourie coefficients:

\begin{equation}
\Delta_{11}=\frac{\sum\limits_{k=1}^{n}(N_{k}-N_{0,k})\cos2\theta_{k}}{\sum\limits_{k=1}^{n}N_{0,k}\cos^{2}2\theta_{k}}
\end{equation}

\begin{equation}
\Delta_{21}=\frac{\sum\limits_{k=1}^{n}(N_{k}-N_{0,k})\sin2\theta_{k}}{\sum\limits_{k=1}^{n}N_{0,k}\sin^{2}2\theta_{k}}
\end{equation}

The probability function has amplitude

\begin{equation}
\Delta_{1}=(\Delta_{11}^{2}+\Delta_{21}^2)^{1/2}
\end{equation}

with the standard deviation of the amplitude

\begin{equation}
\sigma(\Delta_{1})\approx{(\frac{2}{nN_{0}})^{1/2}}.
\end{equation}

The amplitude $\Delta$ was calculated using higher Fourie
coefficients till $4\theta$, according God{\l}owski (1994).

Here we extend our analysis in comparison with Paper II using
autocorrelation and Kolmogorov {\--} Smirnov (K-S) testing.
Autocorrelation test we applied in form:

\begin{equation}
C=\sum\limits_{k=1}^{n}\frac{(N_{k}-N_{0,k})\cdot(N_{k+1}-N_{0,k+1})}{\sqrt{N_{0,k}N_{0,k+1}}}
\end{equation}

Because of the small number of galaxies in some clusters, we made
1000 simulations of the distribution of position angles in 43
fictitious clusters, each cluster with the number of galaxy
members identical to the real cluster. On this basis we obtained
the probability density function (PDF) and the cumulative
distribution function (CDF) seen in Fig. 1 and Fig. 2. The
expected value for the analyzed statistics and its variance were
computed as well. In Table 2 we present the average values of the
analyzed statistics, the corresponding standard deviations, the
standard deviations in the sample, and the standard deviations for
the distribution of $P$ angles. Details of the applied statistics
were presented in previous papers (Paper I and God{\l}owski,
2012).

It is now possible to compare the results obtained for the actual sample
of 43 clusters with that obtained from numerical simulations
(right hand side of Table 2). If we assume that the true
distribution of position angles is uniform, then an exact value for the
probability that the analyzed statistic included a specific
chosen value can be obtained from CDF (Figs. 1 and 2).

However, one should note that our procedure computes the mean
values of the analyzed statistics. When the errors are normally
distributed (Gaussian), which is the case at least for some
statistics analyzed in God{\l}owski (2012), the parameters are
estimated by the maximum-likelihood method. The distribution
should have an asymptotic normal (Gaussian) appearance, which was
checked by God{\l}owski (2012) with the use of  the Kolmogorov
{\--} Lilliefors test (Lilliefors, 1967). There it was shown that
a Gaussian approximation works well, which made the interpretation
of the results much easier. For the sample of all 43 clusters
located in superclusters the distribution of position angles of
galaxy members in the cluster is anisotropic and the departure
from isotropy is usually greater than $3\sigma$ (see Table 2),
with the exception of the first autocorrelation test where the
effect is less than $2 \sigma$. For the angles which gave the
spatial orientation of galaxy planes ($\delta_D$ and $\eta$
angles) the anisotropy is even greater than in the case of
position angles $P$. In our opinion that can be attributed to
incorrectly assumed shapes for the galaxies. That problem was
analyzed in detail by God{\l}owski \& Ostrowski (1999) and
God{\l}owski (2011). Those studies were based on Tully's NGC
Catalogue (1988). In that catalogue, while calculating galaxy
inclination angles, Tully assumed that the ``true'' ratio of axes
for galaxies is 0.2, which, as we have shown in the above papers,
is a rather poor approximation, especially for non-spiral galaxies
(God{\l}owski, 2011). For that reason, the previous study
concentrated on the analysis of position angles. In our present
analysis, presented below, the effect is not especially important
because, for the case of analyzing the spatial orientation of
galaxy planes, our interest is only to show how the alignment
changes with membership of the clusters in a supercluster as well
as with supercluster multiplicity.

The main goal of this study is connected with finding trends
appearing in the data. In Paper I, while analyzing entire samples
of $247$ rich Abell clusters, we found that the alignment
increases with cluster richness. In the analyzed sample of
43 clusters of galaxies belonging to superclusters we do not
observe that effect (Table 3). This conclusion is significantly
different from the result obtained in Paper I for the whole sample of
$247$ rich Abell clusters. We suppose that such a difference can be traced
to environmental effects during the formation o superclusters.
Note that the distributions of analyzed angles are
anisotropic in both cases: for the entire collection of $247$ rich Abell
clusters and for the subsample of $43$ clusters belonging to
superclusters.

In Paper II we presented an analysis of $43$ clusters (Table 4.),
binned according to supercluster multiplicity.
One can observe that the anisotropies seem to decrease with
supercluster richness. For that reason, in the present paper
we decided to perform an unbinned analysis of the linear regression
between values of the analyzed statistics and supercluster
multiplicity. The results are presented in the Table 5. We
analyzed statistics $T= \frac{a}{\sigma(a)}$, the Students' $T$
distribution with $n-2$ degrees of freedom. For $n=46$ at the
significance level $\alpha = 0.05$, the critical value
$T_{cryt}=1.68$. We tested the $H_0$ hypothesis that the value of the analyzed
statistic does not depend on supercluster richness against the
$H_1$ hypothesis that it decreases with supercluster richness.
From Table 5 we can conclude that only in the case of the
$\eta$ angle, the anisotropy decreases with the supercluster
richness and is statistically significant on a significance
level of $0.05$.
\\[2mm]

{\bf 4. Conclusions}\\[1mm]

In the present paper we investigated a sample of $43$ rich Abell
galaxy cluster belonging to a supercluster and containing at least
100 members in the considered area. We found that the orientation
of galaxies in the analyzed cluster was not random. However, in
contrast with the results of God{\l}owski et al. (2010), we detect
that for our sample the alignment of galaxies does not depend on
cluster richness. The differences between samples analyzed in
these studies are as follows. In God{\l}owski et al. (2010), we
analyzed a sample of $243$ rich Abell galaxy clusters, while in
the present paper we analyzed only subsamples of galaxies
belonging to supercluster. Nevertheless, for both samples we
observed that the distributions of analyzed angles $P$, $\delta_D$
and $\eta$, which specify the orientation of galaxies in space,
are not random. We also found that the alignment decreases with
supercluster richness, although the effect is statistically
significant only for azimuthal angles ($\eta$ angles). The results
obtained, which include the dependence of galaxy alignment on
cluster location inside or outside a supercluster as well as
supercluster multiplicity, clearly support the influence of
environmental effects on the origin of galaxy angular momenta. The
problem of obtaining the angular momenta of galaxies in a
structure is rather complicated since several mechanisms play
roles. According to the major scenarios for galaxy formation, in
some cases the angular momentum of galaxies results from local
anisotropic collapse of protostructures, in others due from a
tidal torque mechanism. Moreover, clusters can merge, introducing
additional factors which influence the observed distribution of
galaxy angular momenta. This suggests that environment played a
crucial role in the origin of galaxy angular momentum. In a very
simple and naive picture, if the alignment of galaxies is
primordial, the strongest effect should be observed in small
structures. In the present paper we analyzed only the sample very
rich clusters. For final confirmation or rejection of this
hypothesis, it is necessary to enlarge the analysis taking into
account a sample of poorer clusters . Fortunately, our basic PF
Catalogue (Panko \& Flin, 2006) will allow us to perform such an analysis in the future.
\\[2mm]

{\it Acknowledgements. This research has made use of the NASA
Astrophysics Data System, and was partially supported by grant
BS052 of Jan Kochanowski University (Kielce, Poland).}
\\[3mm]

\noindent
{\bf References\\[2mm]}
\noindent
Abell, G.O., Corwin, H.G., Jr., Olowin, R.P., 1989, {\it ApJS}, {\bf 70}, 1\\
Aryal, B., Saurer, W., 2000, {\it A\&A}, {\bf364}, L97\\
Bower, R. G., Benson, A.J., Malbon, R., Helly, J., Frenk, C. S., Baugh, C. M., Cole, S., Lacey, C. G. 2006, {\it MNRAS}, {\bf370}, 645\\
Catelan, P., Theuns, T. 1996 {\it MNRAS}, {\bf282}, 436\\
Doroshkevich, A. G. 1973, {\it ApL}, {\bf14}, 11\\
Flin, P., God{\l}owski, W. 1986, {\it MNRAS}, {\bf222}, 525\\
Flin, P., Vavilova, I.B., 1996, {\it Proceedings of the 27th
Meeting of the Polish Astronomical Society, held in Poznan,
September 12, 1995}, edited by Marek J. Sarna and Peter B. Marks,
p.63 \\
God{\l}owski, W., 1993, {\it MNRAS}, {\bf265}, 874\\
God{\l}owski, W., 1994, {\it MNRAS}, {\bf271}, 19\\
God{\l}owski, W., 2011, {\it Acta Physica Polonica B}, {\bf 42}, 2323\\
God{\l}owski, W., 2012, {\it ApJ}, {\bf747}, 7\\
God{\l}owski, W., Flin, P., 2010, {\it ApJ}, {\bf708}, 902 \\
God{\l}owski, W., Ostrowski, M., 1999, {\it MNRAS}, {\bf303}, 50
God{\l}owski, W., Panko, E., Flin, P., 2011, {\it Acta Physica Polonica B}, {\bf 42}, 2313, Paper II\\
God{\l}owski, W., Piwowarska, P., Panko, E., Flin, P., 2010, {\it ApJ}, {\bf 723}, 985, Paper I\\
Hawley, D. I., Peebles, P. J. E. 1975, {\it AJ}, {\bf80}, 477 \\
Jones, B., van der Waygaert R., Aragon-Calvo M., 2010 {\it MNRAS}, {\bf408}, 897 \\
Kindl, A. 1987, {\it AJ}, {\bf93}, 1024 \\
Lee, J., Pen, U. 2002, {\it ApJ}, {\bf567}, L111 \\
Li, Li-Xin., 1998, {\it Gen. Rel. Grav.}, {\bf30}, 497 \\
Lilliefors, H. W. 1967, {\it J. Am. Stat. Assoc.}, {\bf62}, 399\\
Navarro, J. F., Abadi, M. G., Steinmetz M. 2004, {\it ApJ}, {\bf613}, L41 \\
Panko, E., 2011, {\it Baltic Astr.}, {\bf20}, 131 \\
Panko, E., Flin, P., 2006, {\it Journ. Astro. Data} {\bf 12}, 1\\
Paz, D.J, Stasyszyn, F., Padilla, N. D. 2008  {\it MNRAS} {\bf389}, 1127 \\
Peebles, P.J.E. 1969, {\it ApJ}, {\bf155}, 393 \\
Shandarin, S.F. 1974, {\it Sov. Astr}. {\bf18}, 392 \\
Silk, J., Efstathiou, G. A. 1983, {\it The Formation of Galaxies, Fundamentals of Cosm. Phys.} {\bf9}, 1 \\
Smargon, A., Mandelbaum, R., Bahcall, N.,  Niederste-Ostholt, M. 2011,  arXiv:1109.6020 \\
Sunyaev, A. R., Zeldovich, Ya. B., 1972 {\it A\&A}, {\bf20}, 189 \\
Trujillo, I., Carretro, C., Patiri, S.G., 2006, {\it ApJ} {\bf640}, L111 \\
Ungruhe, R., Seitter, W., Durbeck, H., 2003, {\it Journ. Astr. Data}, {\bf 9}, 1\\
Varela, J. Rios,J.B., Trujillo, I., 2011, astro-ph 1109.2056 \\
Vavilova, I. B.,  1999, {\it Kinematics Phys. Celest. Bodies},
{\bf 15}, 69\\
Wesson, P. S. 1982, {\it Vistas Astron.}, {\bf26}, 225 \\
Zeldovich, B. Ya. 1970, {\it A\&A}, {\bf5}, 84 \\

\end{document}